\documentstyle[12pt]{article}

\def\be{\begin{equation}}
\def\ee{\end{equation}}
\def\bea{\begin{eqnarray}}
\def\eea{\end{eqnarray}}

\def\nn{\nonumber \\}

\renewcommand{\theequation}{\arabic{section}.\arabic{equation}}

\begin{document}

\begin{titlepage}

\begin{center}
\Large
{\bf Born-Infeld Quantum Condensate as Dark Energy in the Universe}

\vfill

\normalsize

\large{Emilio Elizalde$^\heartsuit$\footnote{Electronic mail: 
 elizalde@ieec.fcr.es}, 
James E. Lidsey$^\diamondsuit$\footnote{Electronic
mail: J.E.Lidsey@qmul.ac.uk}, \\ 
Shin'ichi Nojiri$^\spadesuit$\footnote{Electronic mail: nojiri@nda.ac.jp, 
snojiri@yukawa.kyoto-u.ac.jp} and 
Sergei D. Odintsov$^{\heartsuit\clubsuit}$\footnote{Electronic mail: odintsov@ieec.fcr.es}}

\normalsize

\vfill

{\em $\heartsuit$ 
Consejo Superior de Investigaciones Cientificas, \\
Institut d'Estudis Espacials de Catalunya (IEEC), \\
Edifici Nexus, Gran Capit\`a 2-4, 08034 Barcelona, SPAIN}

\ 

{\em $\diamondsuit$ Astronomy Unit, School of Mathematical 
Sciences,  \\ 
Queen Mary, University of London, Mile End Road, London, E1 4NS, U. K.}

\ 

{\em $\spadesuit$ Department of Applied Physics, 
National Defence Academy, \\
Hashirimizu Yokosuka 239-8686, JAPAN}

\ 

{\em $\clubsuit$ Instituci\`o Catalana de Recerca i Estudis 
Avan\c{c}ats (ICREA), Barcelona, SPAIN \\
and TSPU, Tomsk, RUSSIA}

\end{center}

\vfill

\baselineskip=24pt
\begin{abstract}
\noindent 
Some cosmological implications 
of ultraviolet quantum effects leading to a condensation of 
Born-Infeld matter are considered. It is shown that under 
very general conditions the 
quantum condensate can not act as phantom matter if its energy 
density is positive. On the other hand, 
it behaves as an effective cosmological constant in the limit where 
quantum induced contributions to the energy-momentum tensor dominate 
over the classical effects.

\end{abstract}

\end{titlepage}

\setcounter{equation}{0}

\def\theequation{\arabic{equation}}

Evidence that the universe is 
undergoing a phase of accelerated expansion at the present epoch 
continues to grow. Not only is accelerated dynamics inferred in 
high redshift surveys of type Ia supernovae \cite{Ia}, it 
is now independently implied from observations of the 
anisotropy power spectrum of the Cosmic Microwave Background (CMB)
\cite{wmap,spergel,cmb}. The favoured explanation for 
this behavior is that the universe is presently 
dominated by some form of `dark energy'  contributing
up to 70$\%$ of the critical energy density, with the 
remaining 30$\%$ comprised of clumpy baryonic and non--baryonic dark matter
\cite{spergel}. 
One of the central questions in cosmology today is the origin of this 
exotic matter. In the quintessence scenario, for example, 
the dark energy is a self--interacting scalar field that slowly 
evolves down a potential and thereby acts as a negative pressure source
\cite{quintessence}. 
This paradigm has attracted attention because a wide class 
of models exhibit `tracking' behavior at late times, where
the dynamics of the field becomes independent of its initial 
conditions in the early universe. In principle, 
this may resolve 
the
fine--tuning problem inherent in dark energy models based purely on 
a cosmological constant \cite{tracking}. Nevertheless, there is at 
present no generally accepted origin for the 
quintessence field from a particle physics perspective.

Current observations constrain 
the effective equation of state of the dark energy to be within the region
bounded by  
$-1.45< w_{\rm DE} < -0.74$ at the 95$\%$ confidence 
level \cite{MMOT,spergel}, 
where $w_{\rm DE} \equiv p_{\rm DE}/\rho_{\rm DE}$ and $p$ and 
$\rho$ denote the pressure and energy density of the field, 
respectively. 
Phantom matter  corresponds to the region of parameter space  
$w_{\rm DE} <-1$, where the scalar field has
negative kinetic energy \cite{caldwell,otherphantom}. 
Although matter of this form is presently consistent with observations, 
the origin of such a scalar field with a 
non-conventional kinetic energy is not understood.
On the other hand, it was recently noted that the phantom field 
need not necessarily be a scalar but may in principle 
have vector or tensor degrees of freedom
\cite{GibbonsP}. Moreover, similarities between phantom matter and 
conformal field theory (CFT) were  discussed in Ref. \cite{NOp}.

In view of the above developments, and given the absence  
of a favoured scalar field model, 
it is important to search for alternative candidates 
for the dark energy/phantom
matter whose origin may be found within the context of 
string/M--theory. One of the more 
unusual types of matter that 
is predicted to arise in string/M--theory 
is the so-called Born-Infeld (BI) field.
(For a review, see, e.g., Refs.  \cite{tseytlin}). 
The action for the BI field coupled to gravity is given by 
\be
\label{BI1}
S_{\rm BI}= - \lambda \int d^4 x \left\{
\sqrt{- \det\left(g_{\mu\nu} + F_{\mu\nu}\right)} - 
\sqrt{- \det g_{\mu\nu}} \right\}\ ,
\ee
where $F_{\mu\nu}$ is the field strength for the gauge field, 
$g_{\mu\nu}$ is the spacetime metric 
and $\lambda$ is a coupling constant. 

The question we address in this paper is 
whether such a field can play the role of
dark energy/phantom matter in the present--day universe.   
For simplicity, we assume throughout that the gauge group is abelian. 
It is known that standard abelian BI cosmology is necessarily anisotropic 
(or inhomogeneous) \cite{BI}. On the other hand, as shown 
in Ref. \cite{galtsov}, non--abelian BI cosmology may be isotropic
when the proper choice of gauge field configuration is made.  
Moreover, the equation of state of BI matter may become negative
in some regions of parameter space \cite{galtsov}. 
However, it could be argued that such a complicated choice 
for the gauge field strength might be artificial.
In the present letter, therefore, 
we suggest that 
abelian BI theory may be employed as a toy model for 
isotropic BI 
cosmology where the field strength is time-dependent due to quantum effects
(specifically effects similar to that of gluon condensation in QCD).

In a four-dimensional spacetime, it follows that 
\be
\label{BI2}
\det\left(g_{\mu\nu} + F_{\mu\nu}\right) = (- g)\left\{ 1 + {1 \over 2}
F_{\mu\nu}F^{\mu\nu} - {1 \over 16(-g)}\left(F_{\mu\nu}F^{*\mu\nu}\right)^2
\right\}\ ,
\ee
where $g\equiv \det g_{\mu\nu}$ and $F^{*\mu\nu}\equiv{1 \over 2}
\epsilon^{\mu\nu\rho\sigma}F_{\rho\sigma}$.  
The energy momentum tensor for the BI field is then 
derived by varying the action (\ref{BI1}) with respect to the metric tensor:
\be
\label{BI3}
T_{\rm BI}^{\mu\nu}  \equiv
{1 \over \sqrt{-g}}{\delta S_{\rm BI} \over \delta g_{\mu\nu}}
= - {\lambda \over 2}\left\{ {g^{\mu\nu}\left(1 + {1 \over 2}F_{\rho\sigma}F^{\rho\sigma} 
\right) - F^\mu_{\ \rho}F^{\nu\rho} \over \sqrt{ 1 + {1 \over 2}
F_{\rho\sigma}F^{\rho\sigma} - {1 \over 16(-g)}\left(F_{\rho\sigma}F^{*\rho\sigma}
\right)^2}} - g^{\mu\nu} \right\}\ .
\ee
We now assume that the spacetime metric corresponds to the 
spatially flat, isotropic, Friedmann--Robertson--Walker (FRW) universe:
\begin{equation}
\label{FRWmetric}
ds^2 =-dt^2 +a^2(t) dx^2  ,
\end{equation}
where $a(t)$ represents the scale factor of the universe. 
If the electric or magnetic component of the field strength, $F^{\mu\nu}$, 
becomes non--trivial, the isometry of the metric (\ref{FRWmetric}) 
is broken. 
However, it is known that in the case 
of QCD, the vacuum expectation value of the square of the 
field strength becomes non-trivial due to the condensation of the gluon. 
Phenomenologically, the gluon condensation has been observed by using 
the operator 
product expansion \cite{SVZ}. In QCD the condensation can be derived by 
using the trace anomaly induced effective action \cite{KF} (for 
a
general 
discussion, see \cite{Muta}). 
If we take into account such effects, we may impose as
our main assumption that  
\be
\label{BI4}
\left< F_{\mu\nu}F^{\mu\nu} \right>_V = \alpha   (t) \ ,\quad 
\left< F_{\mu\nu}F^{*\mu\nu}\right>_V = \beta (t) \sqrt{-g}   ,
\ee
where $\left<\ \right>_V$ denotes vacuum expectation 
values and the functions $\alpha(t)$ and $\beta (t)$ may in general depend on 
time but are constant on the spatial hypersurfaces. 
Due to the isometry of the spatial hypersurfaces, we may further assume that 
\be
\label{BI5}
\left< F^0_{\rho}F^{0\rho} \right>_V = {\alpha_t \over 4}g^{00}\ ,\quad 
\left< F^i_{\rho}F^{j\rho} \right>_V = {\alpha_s \over 4}g^{ij}\ ,
\quad \alpha_t + 3\alpha_s =4\alpha   ,
\ee
where $\alpha_s =\alpha_s (t)$, $\alpha_t = \alpha_t (t)$. 

We emphasize that the origin of Eq. (\ref{BI4}) is purely 
quantum in nature and arises from the condensate that appears due to 
vacuum fluctuations.
In general, the vacuum expectation values (\ref{BI4}) 
are not independent. 
However, the relationship between them can only be determined 
by a direct calculation to some finite order in loop corrections (normally 
chosen to be the one--loop level). 
Moreover, such a calculation depends on the choice of 
the background, the origin of the BI field itself, as well as 
the compactification scheme and particular string theory under 
consideration.  
In Ref. \cite{Dbrane}, 
the one-loop effective potential (and static potential) 
for a toroidal D-brane described by the BI-action
 in constant electric and magnetic
fields was evaluated. In the case of the one-loop 
potential, it was
found that the presence of a 
magnetic background may stabilize the D-brane, 
whereas, in contrast, a constant electrical field 
leads to destabilization. The main conclusion to be drawn 
from such a study is that
the consideration of quantum effects in BI theory is extremely involved 
even for the case of a stationary background.
Since the explicit calculation of 
the time--dependence of (\ref{BI4}) is beyond the scope of the present work,   
we adopt a more
phenomenological approach.  
Our aim is to 
discuss the possible role of the quantum 
condensate in cosmological settings and, 
in particular, to determine 
the conditions that $\alpha$ and $\beta$ would 
need to satisfy in order for the BI field to 
act as a viable candidate for dark energy. 

The classical 
thermodynamics of the radiation must also be accounted for. 
The classical contribution to the 
energy density and pressure of the BI field is determined by averaging over 
the spatial volume, as in Ref. \cite{shiromizu}. 
By identifying the electric
and magnetic components such that  
$E_i\equiv F_{i0}$ and $B_i\equiv {1 \over 2}
\epsilon_{ijk} F^{jk}$, respectively, 
we may specify $\left<\sum_i E_i^2\right>
=\left<\sum_i B_i^2\right> \equiv \epsilon (t)$ by invoking the equipartition 
principle \cite{shiromizu}. Since 
$\left< F^0_{\ \rho}F^{0\rho} \right> = \left<\sum_i E_i^2\right>$ and 
$\left< F_{i \rho}F_j^{\ \rho} \right> = - \left<E_i E_j\right> 
+ 2\left<B_i B_j\right>$,  
it is also consistent to 
further assume that $\left<E_i E_j\right>=\left<B_i B_j\right>
={\epsilon}g_{ij} /3$, and in this case, it follows that
\be
\label{BI5b}
\alpha_t = \alpha - 4\epsilon\ ,\quad \alpha_s = \alpha + {4 \over 3}\epsilon\ .
\ee

The parameters $\alpha_t$ and $\alpha_s$ are therefore to be viewed as 
expressing the time--dependence of the averaged  components of the 
field strength, 
$F^0_{\rho}F^{0\rho}$ 
and 
$F^i_{\rho}F^{j\rho}$, respectively, where the contributions from the 
classical radiation bath and the quantum condensate have both been 
taken into account. We may regard the contribution from 
$\alpha$ as arising purely from quantum mechanical effects. 
Since the 
electric field is orthogonal to the magnetic field in the radiation, 
$\sum_i E_i B_i=0$, it also follows that 
$\left< F_{\mu\nu}F^{*\mu\nu}\right> \propto \left< \sum_i E_i B_i
\right> = 0$ at the classical level. This 
corresponds to specifying $\beta =0$. 

We now proceed to determine the effective equation of state for 
the BI field. When   
the connected parts of the operators in the vacuum expectation 
values are neglected, i.e., when 
\be
\label{BI6}
\left< \left(F_{\mu\nu}F^{\mu\nu} \right)^n \left( F_{\mu\nu}F^{*\mu\nu}\right)^m
\right>_V=\alpha^n \beta^m\ ,\quad \mbox{etc.,}
\ee
the expressions for the energy 
density, $\rho_{\rm BI}$, and  pressure, $p_{\rm BI}$, of the BI field follow 
directly from Eq. (\ref{BI3}). We find that 
\bea
\label{BI7}
\rho_{\rm BI} &=& {\lambda \over 2}\left({1 + {\alpha \over 2} - {\alpha_t \over 4} 
\over \sqrt{ 1 + {\alpha \over 2} - {\beta^2 \over 16}}}-1\right)\ ,\\
\label{BI7b}
p_{\rm BI} &=& -{\lambda \over 2}\left({1 + {\alpha \over 2} - {\alpha_s \over 4} 
\over \sqrt{ 1 + {\alpha \over 2} - {\beta^2 \over 16}}}-1 \right)\ .
\eea
Since the BI field is minimally 
coupled to Einstein gravity, its
dynamics is determined 
by the conservation of its energy--momentum, $\nabla_{\mu} T^{\mu\nu}_{\rm BI} 
=0$. For the FRW metric (\ref{FRWmetric}), 
this reduces to the ordinary differential equation:
\begin{equation}
\label{fluid}
\dot{\rho}_{\rm BI} +3H(\rho_{\rm BI} +p_{\rm BI} ) =0  ,
\end{equation}
where $H \equiv \dot{a}/{a}$ 
represents the Hubble expansion parameter and 
a dot denotes differentiation with respect to cosmic time. 
In the purely classical limit, $\alpha =\beta =0$, we find that 
the equation of state corresponds to that of a relativistic fluid, 
$p_{\rm BI} = \rho_{\rm BI} /3$, as expected, and 
this implies that the energy 
density redshifts with the expansion of the universe 
in the standard fashion, $\rho_{\rm BI} \propto \epsilon \propto a^{-4}$.

We now consider the effects of the quantum condensate. 
In general, a matter degree of freedom 
may be viewed as phantom matter if it has positive 
energy density and negative pressure and if
its equation of state satisfies $w \equiv p/\rho < -1$, 
i.e., if $p + \rho > 0$ \cite{caldwell}. This implies that its 
energy density increases with time. By comparing Eqs. (\ref{BI7}) 
and (\ref{BI7b}), we deduce immediately that
\begin{equation}
\label{sum}
\rho_{\rm BI} +p_{\rm BI} = \frac{\lambda}{8} 
\frac{\alpha_s -\alpha_t}{\sqrt{1+\frac{\alpha}{2} -\frac{\beta^2}{16}}}
\end{equation}
and it follows immediately that 
a necessary condition for the quantum BI condensate to 
act as phantom matter is that 
$\alpha_t > \alpha_s$. However, 
since 
$\epsilon$ is a semi--positive--definite quantity,
we conclude from Eq. (\ref{BI5b})  
that $\alpha_t< \alpha_s$ is always satisfied. Thus, the BI field 
can not act as phantom matter in the context discussed here.
This is a general result and is independent of the explicit time--dependence 
of the expectation value of the field strength. 

On the other hand, in the limit where the quantum condensate dominates 
the classical contributions, $\alpha \gg \epsilon$, it follows from 
Eq. (\ref{BI5b}) that $\alpha \approx \alpha_s \approx \alpha_t$, 
and comparison of Eqs. (\ref{BI7}) and (\ref{BI7b}) then 
implies that 
\begin{eqnarray}
\label{BI8}
\rho_{\rm BI} = - p_{\rm BI} = {\lambda \over 2}\left({1 + {\alpha \over 4}  
\over \sqrt{ 1 + {\alpha \over 2} - {\beta^2 \over 16}}}-1\right) \\
\label{BI9}
w_{\rm BI} \equiv {p_{\rm BI} \over \rho_{\rm BI}}= -1\ .
\end{eqnarray}
In other words, in the limit where the classical 
contribution is negligible, the BI field {\em behaves 
precisely as an effective cosmological constant}. This is 
remarkable, given that the initial time dependence of the 
quantum contributions, $\alpha$ and $\beta$, has not 
been specified in the analysis. 
For consistency with Eq. (\ref{fluid}), we require that 
the ratio
\begin{equation}
{1 + {\alpha \over 4}  
\over \sqrt{ 1 + {\alpha \over 2} - {\beta^2 \over 16}}}
\end{equation}
be time--independent and this imposes a restriction on the 
functional form of the parameters $\{\alpha_s , \alpha_t , \beta \}$, 
but does not necessarily imply that they should be time--independent 
themselves. We further require that 
\be
\label{BI11}
{\alpha \over 2} > - 1 + {\beta^2 \over 16} 
\ee
and, in the limit where $8 \alpha \to - 16 + \beta^2$ or 
$\alpha\to + \infty$, both the energy density and pressure diverge. 
It also follows from the identity
\be
\label{BI10}
\left(1 + {\alpha \over 4}\right)^2 - \left( 1 
+ {\alpha \over 2} - {\beta^2 \over 16}\right) = {\alpha^2 + \beta^2 \over 16}\geq 0\ ,
\ee
that the energy density is positive (negative) for 
positive (negative) coupling parameter, $\lambda$. 

The question that now arises, therefore, 
is whether further constraints can be imposed on the 
time--dependent parameters from cosmological considerations.
If the BI condensate is to act as a viable dark energy candidate, 
it can only be starting to dominate the 
energy density of the universe at the present epoch. This 
implies that the parameters $\{ \alpha , \beta \}$ should be evolving in 
such a way that they are much less then unity today, otherwise the coupling 
parameter, $\lambda$, 
would have to be severely fine--tuned. Consequently, 
it is natural to Taylor expand Eqs. (\ref{BI7}) and (\ref{BI7b}) 
to first--order in these parameters and we deduce that the effective equation 
of state in this limit is given by 
\begin{equation}
\label{effective}
w_{\rm BI} \approx  -1 + \frac{128\epsilon}{96\epsilon +3\beta^2}  .
\end{equation}

Eq. (\ref{effective}) illustrates how the classical contribution 
to the field strength, $\epsilon$, pushes the equation of state away from that 
of a cosmological constant. If
the quantum condensate contribution to the energy density 
redshifts more 
slowly than the classical sector as the universe expands, 
the behavior of the 
BI field will gradually approach that of a cosmological 
constant as time proceeds.
Consistency with observations
requires that the second term on the right hand side of Eq. (\ref{effective}) 
should not exceed $0.26$ by the present epoch 
\cite{MMOT,spergel}. It is interesting that the equation of state 
(\ref{effective})
is independent of $\alpha$ at this level of approximation.

To summarize, therefore, we have considered the possible cosmological 
implications of including 
ultraviolet quantum effects on Born--Infeld matter degrees of freedom that 
generate a condensate similar to that of the gluon condensate 
in QCD. 
In a FRW 
background, such effects introduce time--dependent 
corrections to the energy and pressure of the BI field, thereby 
altering its equation of state away from that of a classical radiation 
fluid. Since, in general, the determination of the time--dependence 
of such  
corrections is highly involved, we have invoked a phenomenological 
approach by investigating the conditions 
that should be satisfied by the quantum condensate 
if it is
to act as a viable candidate for dark energy or phantom matter. 
It was found that such a field can not act as phantom matter
unless it has negative 
energy density. Such a no--go result follows even though 
the precise time--dependence of the quantum corrections is unknown. 
On the other hand, we note that although negative energy densities may 
appear to be unphysical, spatial distributions 
of negative energy sources have recently been investigated in detail 
in Ref. \cite{ford}, where it was shown by specific example 
that there exist distributions that are indeed 
allowed\footnote{In our case, we note that 
in the somewhat unlikely event that 
the classical contribution is absent, 
the ansatz $\alpha_t=3\alpha$ and
$\alpha_s=\alpha/3$ could be made and this 
would yield phantom-like behavior for the BI condensate.
However, a definitive conclusion can only 
be drawn after a complete calculation has been performed.}.
Similarly, the consideration of the Casimir energy in various models 
(see, for instance, Ref. 
\cite{elizalde}) shows that quantum corrections to the energy
density may be negative. 

In the classical limit, 
the BI field behaves as a relativistic perfect 
fluid. Surprisingly, however, in the opposite limit 
where the classical contribution is negligible 
relative to the quantum condensate, 
the field acts as 
an effective cosmological constant. There exists an intermediate regime 
of parameter space where the field may act as a dark energy source
under the very weak condition that the classical sector redshifts 
with the expansion of the universe more rapidly than that of the 
quantum condensate. 

This is potentially 
interesting because it suggests a mechanism for 
addressing the fine--tuning problem associated with dark energy cosmology. 
This is the problem of 
understanding why the dark energy has such a low density relative to 
the Planck scale. 
A possible scenario that could be considered 
is the case where the classical sector
dominates the energy density of the BI field 
shortly after the condensate forms
in the very early universe 
and remains the dominant contribution until a relatively recent epoch 
is reached that 
corresponds to a redshift of  
$z \approx 
{\cal{O}} ({\rm few})$. 
In this case, the BI field may act as a relativistic degree of freedom
for most of the history of the universe 
without violating the observational limits imposed by 
primordial nucleosynthesis and CMB constraints. If we further assume
that equipartition holds in the early universe, the energy density of the 
BI field would remain comparable to that of the background CMB
during this time. 
However, if the 
time dependence of the quantum parameters then changed
so that these quantities redshifted more 
slowly than the classical contribution, the BI field would 
transform into 
a cosmological 
constant, but with a sufficiently low energy density at the present 
epoch. It would 
soon come to dominate the universe as its energy density remained
approximately constant. 

In order to proceed further, a number of 
unresolved questions should be addressed. 
In particular, the action (\ref{BI1}) contains 
a string coupling constant within the field strength $F_{\mu\nu}$. It is 
important, therefore, to consider the energy scales associated 
with the condensates that we have considered. 
The coupling constant 
of the BI theory is predicted by string theory 
to be much higher than the observable, present--day dark energy.  
However, the BI theory arises from higher--dimensional 
string theory as an effective $D$--brane theory with an 
associated brane tension, and 
since we have considered the case where the quantum contributions 
$\alpha$ and $\beta$ are of the order unity or less,
the only scale that arises is that of the brane tension. In the standard 
$D$--brane setting, it is expected that  
the bulk cosmological constant and 
brane tension should both be of the order of the Planck scale. 
On the other hand, in warped
compactification models, such as the Randall--Sundrum scenarios 
\cite{rs2}, 
the (higher--dimensional) Planck scale might be considerably 
lower. Moreover, it has been argued 
that 
the confinement of QCD might be dual to the Higgs mechanism
\cite{thooft}. In this 
case, small values for
the parameters $\alpha$ and $\beta$ would arise naturally  
since they would be determined 
by the ratio of the confinement or Higgs scale with 
respect to the four--dimensional Planck scale. Alternatively, it would be 
interesting to consider mechanisms such as renormalization group screening 
to reduce the magnitude of the induced cosmological constant. 
Of course, this very important question of 
physical energy scales can only be addressed concretely 
after a direct
calculation. We propose that 
the scenario we have outlined above provides strong motivation 
for considering 
the field theoretic issues that are involved in 
determining the explicit time--dependence of the 
condensate vacuum expectation 
values in an expanding FRW universe.

Finally, we remark 
that the quantum effects associated with CFT
matter may also be accounted for by
including the contributions due to the 
conformal anomaly. These take the general form 
\be
\label{OVII}
T=b\left(F+{2 \over 3}\Box R\right) + b' G + b''\Box R\ ,
\ee
where $F$ is the square of four--dimensional
Weyl tensor and $G$ is the Gauss-Bonnet invariant. 
If there are  $N$ scalar, $N_{1/2}$ spinor, $N_1$ vector,
$N_{\rm HD}$ higher derivative conformal scalars,  
and $N_2$ ($=0$ or $1$) graviton fields present, then $b$, $b'$ and $b''$, 
are given by
\bea
\label{bs}
&& b={N +6N_{1/2}+12N_1 + 611 N_2 - 8N_{\rm HD} 
\over 120(4\pi)^2}\nn 
&& b'=-{N+11N_{1/2}+62N_1 + 1411 N_2 -28 N_{\rm HD} 
\over 360(4\pi)^2} \nonumber \\
&& b''=0   ,
\eea
respectively.
The contributions  to the energy density and pressure 
due to the conformal anomaly 
have been found explicitly when the metric has the FRW form 
(\ref{FRWmetric}) \cite{NOev}. 
Specifically, for the case of pure de Sitter 
space, it was found that 
\be
\label{phtm2}
\rho_A=-p_A = - {6b'H^4}   .
\ee
It is remarkable, that for higher derivative conformal scalar 
the quantum CFT energy density becomes negative as well.
It then follows that the (nearly de Sitter) Friedmann 
equation is given by 
\be
\label{BI18}
H^2={8\pi \over 3 m^2_P}\left\{{\lambda \over 6}\left({1 + {\alpha \over 4}  
\over \sqrt{ 1 + {\alpha \over 2} - {\beta^2}}}-1\right)
 - 6b' H^4\right\}\ ,
\ee
where $m_P$ is the Planck 
mass and we have neglected any classical effects. 
(For the exact expression
of the CFT energy density in an arbitrary FRW spacetime, see \cite{NOev}).
 
Hence, we deduce that BI quantum effects combined with the 
vacuum polarization which arises due
to CFT matter contributions
also serve as an effective cosmological constant 
with a negative equation of state
and may in principle serve as a mechanism for 
realizing an accelerated phase of cosmic expansion.

\vspace{.3in}
\centerline{\bf Acknowledgments}
\vspace{.3in}
This research is supported in part by the Royal Society (JEL),
Ministry of Education, Science, Sports and Culture of Japan under 
grant number 13135208 (SN) and the 
DGI/SGPI (Spain) project BFM2000-0810 (EE and
SDO).

\end{document}